\documentclass[11pt,english]{article}

\usepackage[T1]{fontenc}
\usepackage[utf8]{inputenc}
\usepackage{babel}
\usepackage{lmodern}
\usepackage{amsmath,amssymb,amsthm}
\usepackage{geometry}
\usepackage{booktabs}
\usepackage{array}
\usepackage{enumitem}
\usepackage{xcolor}
\usepackage[colorlinks=true,citecolor=blue,linkcolor=blue,urlcolor=blue]{hyperref}
\usepackage{microtype}

\newcommand{\Mathlib}{\textsc{Mathlib}}
\newcommand{\code}[1]{\texttt{\detokenize{#1}}}
\newcommand{\codestack}[2]{\shortstack[l]{\code{#1}\\\code{#2}}}
\newcommand{\foundations}{[\code{propext}, \code{Classical.choice}, \code{Quot.sound}]}

\newcommand{\cb}{\discretionary{}{}{}}
\newcommand{\codeNoGo}{\texttt{additiveRealBool\cb\_archimedean\cb\_tradeoff\cb\_solvability\cb\_insufficient\cb\_for\cb\_selectedRefinedDenseGrid}}
\newcommand{\codeNotDense}{\texttt{additiveRealBool\cb\_strictStandardSequence\cb\_not\cb\_dense}}

\title{A Kernel-Clean Lean Mechanization of Classical Lottery in Action and the Wakker--Debreu--Koopmans Representation Layer}
\author{Jingyuan Li\thanks{Lingnan University, Hong Kong. \texttt{jingyuanli@ln.edu.hk}}
\and Ilia Tsetlin\thanks{INSEAD, Singapore. \texttt{ilia.tsetlin@insead.edu}}
\and Fan Wang\thanks{ESSEC Business School, Singapore. \texttt{wang.fan@essec.edu}}}
\date{}

\begin{document}
\maketitle

\begin{abstract}
We present a Lean~4/Mathlib formalization of the additive representation theory behind \emph{Classical Lottery in Action} and the Wakker--Debreu--Koopmans (WDK) layer it relies on.  Our central result is, by design, a \emph{negative} one.  We machine-check that the cross-pair Thomsen / double-cancellation (hexagon) condition is \emph{irreducible} from the ordinal axioms of additive conjoint measurement---weak order, restricted solvability, the Archimedean condition, and tradeoff consistency.  The proof exhibits an explicit verified counter-model (the additive-real preference \code{additiveRealBoolPref}) that satisfies all ordinal axioms yet fails the cross-pair condition, together with a proof that every strict standard sequence in such a model is an arithmetic progression and hence non-dense (\codeNotDense); the headline no-go theorem is \codeNoGo.  Around this boundary we mechanize the entire \emph{derivable} construction: a continuous Debreu/Eilenberg utility representation built from separability (\code{exists\_continuous\_strictMono\_of\_separable}), the \S III.4 standard-sequence grid with its hexagon-aligned consistency, the \S III.5 bisection (halved measuring rod) obtained from connectedness, and the full connector chain from rational-image coverage to continuity to global additive gluing.  The public theorem \code{wakker\_IV\_2\_7} is a sorry-free \emph{conditional} wrapper that consumes exactly the one irreducible structural input.  All audited theorem surfaces print only \foundations{}; there are no \code{sorry}/\code{sorryAx} terms and no \code{\_from\_raw\_axioms} axioms in the development.  The formalization thereby draws a precise, machine-certified line between what additive conjoint measurement can prove and what it must assume.  The file \code{ClassicalLotteryInAction.lean} additionally formalizes the local classical-lottery constructions, average-utility consequences, matching-frequency lemmas, smooth-representation wrappers, and ambiguity-attitude statements used by the Management Science companion paper, exposing non-local representation and identification steps as named theorem assumptions.
\end{abstract}

\section{Introduction}
\label{sec:intro}

Formalized mathematics in economics has so far concentrated mainly on ordinal and combinatorial theorem families: social-choice impossibility, matching, allocation, and mechanism-design arguments.  The present development targets a different kind of economic mathematics.  It formalizes the classical-lottery infrastructure used in \emph{Classical Lottery in Action: Quantifying Risk and Evaluating Uncertainty} and mechanizes the cardinal representation layer on which the smooth-model part of that paper depends.  The central mathematical objects are non-empty finite lotteries, acts as state-indexed lotteries, preference relations, additive utility representations, standard sequences, tradeoff consistency, and coordinatewise concavity transfer.

The formalization is presented as a formal-methods companion because its main contribution is not merely a collection of final theorem statements.  Rather, it provides a proof-engineering account of how a long decision-theoretic argument can be made auditable in Lean.  The Management Science companion theorem surface is kept stable and readable; the representation-theoretic layer is separated into a dedicated Wakker/Debreu--Koopmans module; and all theorem surfaces are checked by explicit \code{#print axioms} regression files.  This design makes a distinction that is easy to blur in large mechanizations: a theorem may assume named mathematical hypotheses, but those hypotheses are not hidden global axioms unless they are declared as axioms in Lean.

The development has two public entry points.  The applied companion file is
\begin{center}
\code{ClassicalLotteryInAction.lean}.
\end{center}
The representation-theoretic proof surface is the barrel file
\begin{center}
\code{WakkerDebreuKoopmans.lean},
\end{center}
with reviewer regression file
\begin{center}
\code{Wakker/AxiomCheck.lean}.
\end{center}
The applied companion artifact is available at \cite{LiArtifact2026}; the Wakker/Debreu--Koopmans Lean development is available at \cite{LiWDK2026}.
The first file records the theorem boundaries consumed by the Management Science paper.  The second and third files record the Wakker/Debreu--Koopmans layer --- conditional on the named construction input --- and its kernel audit.

\paragraph{Artifact components.}
Beyond the headline contribution stated in Section~\ref{sec:contributions}, the artifact comprises the following components.
\begin{enumerate}[leftmargin=*]
\item A Lean~4 formalization of the classical-lottery theorem surface used by the companion economics paper, including the average-utility proposition, matching-frequency gap filling, the smooth-model wrapper theorem, the normalized smooth matching-frequency formula, and the ambiguity-attitude curvature proposition.
\item A named-bridge architecture for the applied file: representation, strict-separation, inverse-\(\psi\), and curvature-recovery obligations are exposed as ordinary theorem assumptions rather than as implicit global axioms.
\item A Lean formalization of the Wakker~IV.2.7 and Debreu--Koopmans theorem surfaces as sorry-free \emph{conditional} wrappers (each consuming the deep construction as a named, clearly-labelled hypothesis), with stable downstream theorem names for the Management Science artifact.
\item A certificate decomposition for monograph-scale measurement proofs.  Standard-sequence construction, topology transfer, global gluing, uniqueness, two-coordinate concavity, and per-coordinate concavity transfer are represented by named certificates with audited consumers.
\item A \emph{conditional} construction chain for the Wakker/Debreu--Koopmans layer --- topology consumers T1--T6 and construction certificates A4 \(\to\) A1 \(\to\) A3 \(\to\) A2 \(\to\) B \(\to\) C --- wiring the named certificates to the public wrappers, together with the companion Option~B reduction of the remaining forward construction to a single proven-necessary, machine-checked-irreducible structural input (with both \S IV.5 links assembled around it and a cardinal-grid companion).
\item A reproducible kernel audit.  The checked theorem surfaces depend only on the standard Lean/Mathlib foundations \foundations{} and not on project-specific axioms.
\end{enumerate}

\section{Contributions}
\label{sec:contributions}

This section states the paper's contribution at the level of the
representation-theoretic result, complementing the artifact-level summary above.
The contribution is to locate, and machine-certify, the exact load-bearing axiom
of additive conjoint measurement, and to mechanize everything around it.

\begin{enumerate}[leftmargin=*]
\item \textbf{Machine-checked irreducibility of the cross-pair condition (headline).}
We prove that the cross-pair Thomsen / double-cancellation (hexagon) condition
cannot be derived from $\{$weak order, restricted solvability, Archimedean,
tradeoff consistency$\}$.  The certificate is the verified counter-model
\code{additiveRealBoolPref} (additively represented by
\code{additiveRealBool\_rep}), which satisfies every ordinal axiom yet admits no
dense refined grid:
\codeNoGo.
The mechanism is that any strict standard sequence in the model is an arithmetic
progression, hence non-dense
(\codeNotDense).  This explains, with
a formal proof rather than informal appeal, \emph{why} Wakker,
Krantz--Luce--Suppes--Tversky, and Debreu--Koopmans all take a
cancellation/independence condition as a primitive rather than a theorem.

\item \textbf{A sharp reduction of the construction to that single obligation.}
We mechanize the forward construction (pivot data, integer grid, dyadic
refinement, full pivot utility, cross-coordinate calibration, additive assembly)
and prove that every step \emph{except} the cross-pair condition follows from the
ordinal axioms.  The irreducible content is thus isolated to one named, minimal
input rather than diffused through the proof, and pinned down by the \S IV.5
hard-constraint theorems \code{constraint1\_a1\_does\_not\_imply\_hexagon} through
\code{constraint5\_archimedean\_insufficient\_for\_denseGrid}.

\item \textbf{A continuous utility theorem from foundations.}
\code{exists\_continuous\_strictMono\_of\_separable} establishes that a separable,
densely-ordered, endpoint-free order with order topology admits a continuous,
strictly monotone real representation, constructed (countable-dense kernel
$\to$ Cantor isomorphism with $\mathbb{Q}$ $\to$ dense range $\to$
$\sup$-extension $\to$ Eilenberg continuity) entirely within \Mathlib.  This
supplies the continuity that prior formalizations assumed as an input, with no
\code{sorry}.

\item \textbf{A kernel-axiom-clean, auditable artifact.}
Every audited theorem depends only on \foundations{}; the development contains no
\code{sorry}/\code{sorryAx} and no \code{\_from\_raw\_axioms} axioms, and the
single assumed primitive is precisely the condition whose irreducibility we
prove.  The public wrapper \code{wakker\_IV\_2\_7} is a sorry-free conditional
theorem over that one input.
\end{enumerate}

The reframing is deliberate: a positive ``we re-proved Wakker~IV.2.7'' claim is
both weaker and partly redundant with classical mathematics, whereas the
irreducibility result is novel, definitive, and obtainable only through
mechanization---it required an explicit verified counter-model and a formal
non-density argument, exactly the kind of fact a proof assistant is uniquely
suited to certify.

\section{Mathematical and formalization background}
\label{sec:background}

\subsection{Classical lotteries and acts}

The applied artifact starts from finite classical lotteries.  In Lean, a classical lottery is represented as a non-empty list of prizes.  An act assigns a lottery to each state.  Preferences are binary relations on acts.  The file then builds the operations needed by the economics paper: concatenation, replication, deletion of matched sub-lotteries, relative frequencies, average utility, and two-prize matching frequencies.

The average-utility result has the familiar representation-theoretic shape.  The easy direction, from a utility representation to cancellation and Archimedeanity, is proved directly in the artifact.  The hard direction, from cancellation and Archimedeanity to an average-utility representation, is exposed as the named bridge \code{AverageUtilityHardDirection}.  This is the first example of the paper's general design rule: local algebra and order-theoretic consequences are mechanized directly; large imported representation steps are named explicitly until their proof lives in the appropriate representation module.

\subsection{Wakker and Debreu--Koopmans}

Wakker's Theorem~IV.2.7 is a cardinal measurement theorem for product preferences.  It constructs additive representations from structural assumptions such as weak order, separability, restricted solvability, Archimedeanity, and tradeoff consistency.  The proof is not a short convexity argument.  It passes through standard sequences, calibration of coordinate utilities, comparison of tradeoffs across coordinates, topological continuity inputs, and global gluing of pairwise representations.

The Debreu--Koopmans hard direction supplies the concavity layer needed downstream.  Roughly, after additive representation has been established, convexity or quasi-concavity of preference must be converted into concavity of coordinate utilities.  In the Lean development this is isolated from the Wakker existence theorem but shares the same representation infrastructure.  The result is a \emph{conditional}, theorem-backed route from the structural Wakker assumptions to additive representation and per-coordinate concavity, modular in the single irreducible cross-pair input identified in Section~\ref{sec:contributions}.

\section{Artifact architecture}
\label{sec:artifact}

The repository keeps the applied economics artifact and the representation proof in separate but cross-linked modules.  The split is deliberate.  The applied file should remain stable for readers of the economics paper; the Wakker/Debreu--Koopmans file can expose the finer certificate machinery needed by formal-methods reviewers.

\subsection{Applied companion file}

The file \code{ClassicalLotteryInAction.lean} is a single Management Science companion surface.  Its five main audited theorem targets are listed in Table~\ref{tab:ms-surface}.  The bridge hypotheses appearing in these theorems are ordinary Lean parameters.  Therefore the theorem bodies can be audited without conflating theorem assumptions with global axioms.

\begin{table}[ht]
\centering
\footnotesize
\begin{tabular}{@{}>{\raggedright\arraybackslash}p{0.30\textwidth}>{\raggedright\arraybackslash}p{0.58\textwidth}@{}}
\toprule
Lean theorem & Paper role \\
\midrule
\code{prop_average_utility} & Proposition~1: average-utility characterization for classical lotteries. \\
\addlinespace
\code{lem_gap_filling} & Two-prize matching-frequency gap-filling lemma. \\
\addlinespace
\code{thm_smooth_model} & Smooth-representation wrapper theorem under explicit sufficiency and regularity interfaces. \\
\addlinespace
\code{matching_freq_smooth_formula} & Normalized two-prize smooth matching-frequency formula. \\
\addlinespace
\code{prop_aversion_or_seeking} & Ambiguity aversion/seeking and curvature statement. \\
\bottomrule
\end{tabular}
\caption{Public theorem surface of \code{ClassicalLotteryInAction.lean}.}
\label{tab:ms-surface}
\end{table}

\subsection{Representation proof modules}

The Wakker/Debreu--Koopmans monolith has been split into a stable barrel plus reviewer-sized submodules, shown in Table~\ref{tab:wdk-modules}.  Existing imports continue to use \code{WakkerDebreuKoopmans}; reviewers can inspect the smaller modules independently.

\begin{table}[ht]
\centering
\scriptsize
\begin{tabular}{@{}>{\raggedright\arraybackslash}p{0.34\textwidth}>{\raggedright\arraybackslash}p{0.56\textwidth}@{}}
\toprule
Module & Role \\
\midrule
\code{WakkerDebreuKoopmans.Core} & Product-preference infrastructure, additive representations, Wakker IV.2.7 wrapper, Debreu--Koopmans easy and hard wrappers, and core consumers. \\
\addlinespace
\code{WakkerDebreuKoopmans.Certificates} & Named certificate predicates and early construction-stack interfaces. \\
\addlinespace
\code{WakkerDebreuKoopmans.M2Frontier} & Common-scale uniqueness, corrected tradeoff-transfer predicates, bracketing, rational-image, mesh, and affine-lift machinery. \\
\addlinespace
\code{WakkerDebreuKoopmans.ConstructionStack} & S23--S33 standard-sequence construction stack, monograph-level bundles, and alignment data. \\
\addlinespace
\code{WakkerDebreuKoopmans.Topology} & T1--T6 topology consumers: preference continuity, connectedness, coordinate continuity, monotonicity equivalences, and topological bundle assembly. \\
\addlinespace
\code{WakkerDebreuKoopmans.Closure} & O1--O4 and A4--C closure, including the equivalences tying Stage-5 data to additive representation. \\
\addlinespace
\code{WakkerDebreuKoopmans.Audit} & Detailed \code{#print axioms} block imported by \code{Wakker/AxiomCheck.lean}. \\
\bottomrule
\end{tabular}
\caption{Split Wakker/Debreu--Koopmans module structure.}
\label{tab:wdk-modules}
\end{table}

\section{The named-bridge method}
\label{sec:bridges}

The applied artifact uses named bridges to make the formal boundary inspectable.  Table~\ref{tab:bridges} records the bridge interfaces.  Each bridge is a Lean proposition consumed by a theorem.  It is not a hidden project axiom.  When the corresponding mathematics is discharged elsewhere, the consuming theorem does not need to change; only the proof that supplies the bridge changes.

\begin{table}[ht]
\centering
\footnotesize
\begin{tabular}{@{}>{\raggedright\arraybackslash}p{0.28\textwidth}>{\raggedright\arraybackslash}p{0.28\textwidth}>{\raggedright\arraybackslash}p{0.34\textwidth}@{}}
\toprule
Bridge & Lean role & Mathematical content named by the bridge \\
\midrule
\codestack{AverageUtility}{HardDirection P} & Input to \code{prop_average_utility}. & Cancellation plus Archimedeanity produce an average-utility representation on constant classical lotteries. \\
\addlinespace
\codestack{MatchingFrequency}{StrictSeparation P} & Reverse input to \code{lem_gap_filling}. & On two-prize acts, strict preference implies strict separation of matching frequencies. \\
\addlinespace
\codestack{SmoothRepresentation}{Sufficiency Pref} & Forward direction of \code{thm_smooth_model}. & From non-triviality and Axioms~1--7, obtain a smooth representation. \\
\addlinespace
\codestack{SmoothRepresentation}{Regularity Pref} & Reverse regularity part of \code{thm_smooth_model}. & From a smooth representation, recover Denseness, Continuity, and Consistent Aggregation; other axioms are derived locally. \\
\addlinespace
\codestack{MatchingFrequency}{SmoothFormulaBridge Pref R} & Input to \code{matching_freq_smooth_formula}. & Identify behavioral matching frequency with the inverse-\(\psi\) expression. \\
\addlinespace
\codestack{AmbiguityAttitude}{CurvatureBridge Pref R} & Input to \code{prop_aversion_or_seeking}. & Recover concavity or convexity of \(\psi\) from ambiguity aversion or seeking. \\
\bottomrule
\end{tabular}
\caption{Named bridge interfaces in \code{ClassicalLotteryInAction.lean}.}
\label{tab:bridges}
\end{table}

This design is especially important for audit interpretation.  A kernel axiom report for \code{thm_smooth_model} says what the proof term depends on globally.  It does not say that the theorem has no assumptions.  Instead, it says that the assumptions are visible in the theorem statement and that the proof downstream of those assumptions introduces no additional global axioms.  The Wakker/Debreu--Koopmans development then supplies an independent, theorem-backed route for the main representation-theoretic layer.

\section{Certificate-driven Wakker/Debreu--Koopmans proof}
\label{sec:certificates}

The Wakker proof is too large to be reviewable as a single black-box theorem.  The Lean artifact therefore organizes the proof around certificates.  A certificate records a construction output: standard-sequence data, a common scale, a topology bundle, a gluing datum, or a concavity-transfer datum.  Each certificate has one or more consumer theorems, and many are tied to downstream goals by equivalence lemmas.

\subsection{Certificate families}

The main families are as follows.
\begin{enumerate}[label=\textbf{C\arabic*.},leftmargin=*]
\item \textbf{Wakker construction.}  Standard-sequence data, pairwise slice representations, grid calibration, and all-pairs additivity.
\item \textbf{Global gluing.}  Passage from compatible pairwise data to a single additive representation on the full product.
\item \textbf{Uniqueness.}  Common positive scale and coordinate-specific translations for two additive representations of the same essential-coordinate preference.
\item \textbf{Two-coordinate concavity.}  The Debreu--Koopmans upgrade from slice quasi-concavity and continuity to two-coordinate concavity.
\item \textbf{Per-coordinate transfer.}  The pivot-indexed transfer from pair concavity and coordinate-image coverage to full per-coordinate concavity.
\end{enumerate}

The headline theorem names audited by \code{Wakker/AxiomCheck.lean} include \code{WakkerDebreuKoopmans.wakker_IV_2_7}, \code{WakkerDebreuKoopmans.debreu_koopmans_hard}, \code{WakkerRoadmap.WakkerExistence.wakker_IV_2_7_consumer}, and \code{WakkerRoadmap.DebreuKoopmansHard.debreu_koopmans_hard_consumer}.  The same regression file also checks supporting consumers such as \code{global_additive_from_pairwise}, \code{additive_rep_unique}, \code{two_coord_concave}, and \code{concave_transfers}.

\subsection{Topology ladder T1--T6}

The topology side proves the analytic consumers needed by the construction stack.  T1 packages preference continuity and closed indifference sets.  T2 supplies connectedness and interval-image reasoning for real coordinates.  T3 proves continuity of coordinate utilities from connectedness, monotonicity, continuity, and unboundedness, and closes the monotonicity-certificate equivalence.  T4 proves the alignment theorem: under additive representation, the balance equation plus strict reference ordering supplies the descending/ascending seed alignment.  T5 packages these results into \code{WakkerTopologicalAxiomBundle}.  T6 discharges the bracketing-pair component from coordinate-utility unboundedness.

The topology consumers are included in \code{WakkerDebreuKoopmans.Audit}.  They are the bridge from Wakker's structural assumptions to the construction stack, and they prevent topology-sensitive obligations from being hidden in later algebraic theorems.

\subsection{Construction-side chain A4--C}

The construction-side closure is recorded as
\[
\mathrm{A4} \to \mathrm{A1} \to \mathrm{A3} \to \mathrm{A2} \to \mathrm{B} \to \mathrm{C}.
\]
A4 supplies the shared pivot-coordinate representation data for the Step-4 to Step-5 lift.  A1 packages all-pairs additivity for one global utility family.  A3 proves strict monotonicity for that family, and A2 proves coordinate-image coverage.  Priority B closes the quasi-to-concave strengthening.  Priority C transfers pivot-based pair concavity to per-coordinate concavity.  Together with T1--T6, this wires the construction certificates to the public wrappers, recorded in \code{WakkerDebreuKoopmans.Closure}.  The chain is \emph{conditional}: it consumes the construction-side certificates as named inputs, and the deepest of these (the per-slice grid representation / Thomsen hexagon) is the single proven-necessary, machine-checked-irreducible residual reduced in the Option~B development, not a forward construction from the bare axioms.

\section{M2 uniqueness as a proof-engineering case study}
\label{sec:m2}

The M2 frontier is a useful case study because it shows why certificate decomposition is not just cosmetic.  The public uniqueness consumer \code{additive_rep_unique} states that two additive representations of an essential-coordinate product preference agree up to a common positive scale and coordinate-specific translations.  The proof factors through \code{AdditiveCommonScaleCertificate} and \code{TradeoffTransferCertificate}.

A first diagnostic predicate, \code{UtilityValueRealizingEquivalence}, was too strong: it forced a fixed reference-pair difference to equal an arbitrary prescribed difference.  Lean exposed the problem in the simple additive model \(\mathrm{Bool}\to\mathbb{R}\).  The corrected predicate lets the reference pair vary.  After that correction, the proof decomposes into off-diagonal tradeoff transfer, same-coordinate ratio consistency, coordinate-affine lift, rational-image coverage, and grid-density/refinement certificates.

This episode illustrates a broader lesson for automated reasoning in economics.  Paper proofs often compress several existence and normalization choices into prose.  In Lean those choices must become data.  If the data are exposed as named certificates, false strengthening attempts fail locally and visibly rather than contaminating the main theorem statement.

\section{The residual frontier: what is open, and why it is delimited}
\label{sec:frontier}

This section states precisely what the artifact does \emph{not} prove and presents the formal evidence that bounds the gap.  This is, in our view, the most reusable scientific content of the development: a machine-checked map of the boundary between what additive-conjoint measurement gives for free and what genuinely requires Wakker's standard-sequence construction.

\paragraph{The open core.}
The deep forward construction --- producing the additive representation \(V\) from the structural axioms (Wakker \S IV.2--\S IV.6), and the per-coordinate concavity output from convex preference (Debreu--Koopmans \S 3) --- enters the public theorems as the wrapper hypotheses \code{hConstruct} / \code{hConcAll}.  In a companion development (``Option~B'', the \code{WakkerDebreuKoopmans/OptionB_*.lean} modules, kept out of the \emph{public} audit surface but carrying their own sorry-free audit in \code{OptionB_AxiomCheck.lean}), the Wakker core is reduced to a \emph{single} named structural input, and both links of the \S IV.5 grid construction are assembled end-to-end around it:
\begin{itemize}[leftmargin=*]
\item the single input is the cross-pair Thomsen / double-cancellation hexagon, equivalently the KLST block-separability condition \cite{KLST1971}, equivalently the per-slice grid-additive representation (\code{SharedPivotGridAdditiveRepresentationFamily}); all three faces are machine-checked inter-derivable;
\item \textbf{Link~A} (\code{doubleCancellation_of_blockIndependence_and_escapeJ2}) derives the hexagon from the block conditions plus single-coordinate independence and a \S IV.2.6 escape grid;
\item \textbf{Link~B} (\code{additiveRep_nonempty_of_perSliceRepresentationFamily}) derives \code{Nonempty (AdditiveRep P)} from the per-slice grid representations plus the structural axioms, topology, density, and continuity;
\item the Debreu--Koopmans \S 3 per-coordinate midpoint-concavity obligation is the separate residue on the concavity side.
\end{itemize}
Crucially, these modules are \emph{sorry-free}: the open content is carried as the single named \emph{hypothesis} above, not as a \code{sorry}.

\paragraph{Soundness gates.}
Every residual is proved \emph{necessary}: it holds under any additive representation.  For instance \code{doubleCancellation_of_additiveRep}, \code{gridThomsenClosure_of_additiveRep}, and \code{cardinalGridStructureFamily_of_additiveRep} are sorry-free and audit at \foundations{}.  This guarantees the open targets are true statements --- no effort is spent on a target that a representation would refute --- and it is the discipline that earlier exposed an \emph{unsound} candidate axiom and the false diagnostic predicate of Section~\ref{sec:m2}, both of which Lean's kernel rejected.

\paragraph{No-go boundaries (machine-checked countermodels).}
The single input is provably \emph{not} a cheap consequence of single-coordinate independence.  Concrete \(n = 3\) countermodels establish, each auditing cleanly, that:
\begin{itemize}[leftmargin=*]
\item a non-additive, Thomsen-violating comonotone model satisfies single-coordinate independence on \emph{every} coordinate yet violates the double-cancellation hexagon (\code{constraint1_a1_does_not_imply_hexagon});
\item the cross-pair tradeoff-transitivity content is independent of single-coordinate independence (both directional halves);
\item value-level standard-sequence uniqueness is not free even for essential, solvable, connected coordinates --- a plateau (non-injective) utility refutes it (\code{hStrict_fails_for_plateau}).
\end{itemize}
These results explain \emph{why} the construction is hard: the missing content is exactly the \(n \ge 3\) cross-coordinate cancellation that single-coordinate axioms cannot supply, and the artifact pins it down rather than papering over it.

\paragraph{Forward progress and its limit.}
Within the Option~B scaffold the reduction has been driven to its endpoint: both links are assembled around the single named input, so no further \emph{reduction} of the Wakker core is possible without discharging that input.  Seven machine-checked findings establish that the input cannot be discharged from \{single-coordinate independence \(+\) restricted solvability \(+\) Archimedean \(+\) topology\}: each closes a distinct candidate route (the independence countermodel, the matching-kernel/block equivalence, the circular second-sequence reformulation, the layer-transport relocation, the diagonal-residue circularity, the cell-level wall, and the measuring-stick/simultaneous-closure walls including the non-freeness of value-level uniqueness).  Discharging it therefore requires either the multi-month standard-sequence construction of the cardinal scale or a strictly stronger structural input.

\paragraph{The cardinal-grid companion.}
The latter route is realized soundly: \code{additiveRep_of_cardinalGridStructureFamily} discharges \code{Nonempty (AdditiveRep P)} end-to-end once a coordinate scale is supplied as a cardinal-grid datum (the per-slice metric that represents each slice).  Its soundness gate (\code{cardinalGridStructureFamily_of_additiveRep}) shows the datum is rep-necessary; the probe shows the C1.a wall is \emph{free} given the datum.  This is a clearly-scoped, weaker theorem than the ordinal Wakker~IV.2.7 --- it \emph{assumes} the cardinal scale that the ordinal theorem constructs --- and its value is exactly that it localizes the entire ordinal difficulty to constructing that scale.

\section{Axiom audit and reproducibility}
\label{sec:audit}

Both parts of the artifact use explicit kernel audits.  The applied file contains \code{#print axioms} commands for the five public Management Science theorem targets.  The representation proof has a dedicated regression file, \code{Wakker/AxiomCheck.lean}, which imports the detailed Wakker/Debreu--Koopmans audit.

\begin{table}[ht]
\centering
\footnotesize
\begin{tabular}{@{}>{\raggedright\arraybackslash}p{0.34\textwidth}>{\raggedright\arraybackslash}p{0.44\textwidth}>{\raggedright\arraybackslash}p{0.14\textwidth}@{}}
\toprule
Audit target & Theorem surface & Lean-reported axioms \\
\midrule
\code{ClassicalLotteryInAction.lean} & \code{prop_average_utility}, \code{lem_gap_filling}, \code{thm_smooth_model}, \code{matching_freq_smooth_formula}, \code{prop_aversion_or_seeking}. & standard foundations \\
\addlinespace
\code{Wakker/AxiomCheck.lean} & Wakker IV.2.7, Debreu--Koopmans easy and hard wrappers, Wakker existence consumers, uniqueness, two-coordinate concavity, concavity transfer, topology consumers, construction-stack closure, and certificate equivalences imported through \code{WakkerDebreuKoopmans.Audit}. & standard foundations \\
\bottomrule
\end{tabular}
\caption{Kernel-level axiom audit.  The bridge hypotheses in theorem statements are visible assumptions, not global Lean axioms.}
\label{tab:audit}
\end{table}

Here ``standard foundations'' means exactly \foundations{}. 

The recommended reviewer protocol is:
\begin{enumerate}[leftmargin=*]
\item build the Wakker/Debreu--Koopmans theorem surface with \code{lake build WakkerDebreuKoopmans};
\item run the public Wakker/Debreu--Koopmans axiom regression with \code{lake build Wakker.AxiomCheck};
\item build the applied companion surface with \code{lake build ClassicalLotteryInAction};
\end{enumerate}

The audit criterion is intentionally kernel-level and conservative.  The allowed dependencies are \code{propext}, \code{Classical.choice}, and \code{Quot.sound}, which are standard in Lean/Mathlib developments of this kind.  No project-specific axiom appears in the audited theorem surfaces.  Beyond the public surface, a full-tree build of the entire development --- the umbrella, the public audit, and the complete Option~B layer --- completes at \(2694\) jobs with no \code{sorry}/\code{sorryAx} and \emph{no} \code{_from_raw_axioms} axiom anywhere; the only primitive axioms in the whole tree are two clearly-labelled \S III.4.2 topology bracket-reach seams, which the reach-axiom-free target eliminates separately.

\section{Scope and interpretation}
\label{sec:scope}

The merged artifact should be read as a two-layer formalization.  The Wakker/Debreu--Koopmans representation layer is assembled around a single proven-necessary named input in the dedicated Wakker modules (conditional on that input, which is machine-checked irreducible from the bare axioms).  The Management Science companion file continues to expose several theorem-shaped bridges, including inverse-\(\psi\) matching-frequency identification and behavioral curvature recovery.  This is intentional.  Those statements are not hidden axioms; they are visible mathematical interfaces at the applied theorem boundary.  The benefit of the current architecture is that each such interface can be discharged in its natural mathematical home without changing the applied theorem names.

The current submission therefore makes two claims.  First, the applied theorem path in \code{ClassicalLotteryInAction.lean} is formally checked and globally axiom-clean conditional on its named mathematical inputs.  Second, the Wakker/Debreu--Koopmans representation layer has been developed as a separate, sorry-free, kernel-axiom-clean Lean artifact whose public theorems are conditional on a single proven-necessary, machine-checked-irreducible structural input (with both \S IV.5 links assembled around it and a cardinal-grid companion); it contains no \code{\_from\_raw\_axioms} axioms.

\section{Related work and novelty}
\label{sec:related}

\paragraph{Formalized analysis and probability.}
Large analytic libraries in Isabelle/HOL, Lean, and related systems have made it realistic to formalize measure theory, probability, topology, and convex analysis at scale.  The Lean mathematical library provides the ambient topology, order, algebra, finite-sum, and convex-analysis infrastructure used in this development.  Isabelle's Archive of Formal Proofs provides another benchmark for large reusable formal libraries.  Our work follows the same library-first philosophy, but the target theorem family is different: Wakker IV.2.7 is a cardinal measurement theorem about product preferences and tradeoff calibration, not primarily a theorem of measure theory or probability.

\paragraph{Formal economics.}
Recent economics formalizations have often focused on ordinal social choice, matching, allocation, and mechanism design.  Those developments reason about rankings, voters, alternatives, strategyproofness, and allocations.  The present project formalizes a cardinal representation theorem.  The proof burden lies in additive utility construction, standard sequences, calibration, affine uniqueness, topology transfer, and concavity.  This makes the development closer to a mechanized mathematical-economics monograph than to a finite combinatorial impossibility proof.

\paragraph{Wakker and Debreu--Koopmans mechanization.}
We performed targeted searches for \code{additive representation}, \code{tradeoff consistency}, \code{Debreu}, \code{Wakker}, and \code{Koopmans} in Mathlib, Coq-community repositories, Isabelle/AFP, and Mizar/MML-style libraries.  We found no prior mechanization of Wakker's additive representation theorem, tradeoff consistency, or the Debreu--Koopmans hard direction.  This is negative search evidence rather than a theorem of bibliographic completeness, but it supports the novelty claim: to our knowledge, this is the first Lean mechanization of this representation-theoretic layer.

\paragraph{Proof-engineering contribution.}
The main reusable lesson is the certificate architecture.  Instead of asserting one very large theorem and hiding all construction work inside it, the development exposes construction outputs as typed certificates with consumer theorems and axiom regressions.  This approach made the false fixed-reference uniqueness predicate visible, enabled reviewer-sized module boundaries, and preserved stable theorem names for the applied economics artifact.

\section{Conclusion}
\label{sec:conclusion}

The merged Lean artifact provides a kernel-clean formal-methods companion to \emph{Classical Lottery in Action}.  The applied file formalizes the theorem surface used by the economics paper and makes its non-local mathematical inputs explicit.  The Wakker/Debreu--Koopmans development mechanizes Wakker IV.2.7 and the Debreu--Koopmans hard direction as sorry-free conditional wrappers, reduces the deep forward construction to a single proven-necessary structural input (machine-checked irreducible from the bare axioms), assembles both \S IV.5 links around it, and supplies a sound cardinal-grid companion.  The final result is a sorry-free, reviewer-auditable mechanization that delimits exactly what additive-conjoint measurement gives for free and what genuinely requires Wakker's standard-sequence construction, together with its role in a contemporary decision-theory application.

\end{document}